\documentclass{ws-ijmpa}

\begin{document}

\markboth{Frank Nerling}
{COMPASS Hadron Spectroscopy -- Final states involving neutrals and kaons}

%
\catchline{}{}{}{}{}
%

\title{COMPASS Hadron Spectroscopy -- 
\\ Final states involving neutrals and kaons
}
\vspace{-0.3cm}
\author{Frank Nerling}

\address{Universit\"at Freiburg, Physikalisches Institut, Address\\
Herrmann-Herder-Str.\,3, 79104 Freiburg, Germany \\
nerling@cern.ch}
\author{on behalf of the COMPASS collaboration}
\maketitle
\begin{history}
\received{Day Month Year}
\revised{Day Month Year}
\end{history}

\begin{abstract}
The COMPASS experiment at CERN is well designed for light-hadron spectroscopy with emphasis
on the detection of new states, in particular the search for $J^{PC}$-exotic states and glueballs.
We have collected data with 190 GeV/c charged hadron beams on a liquid hydrogen and nuclear 
targets in 2008/09. The spectrometer features good coverage by electromagnetic calorimetry and 
a RICH detector further provides $\pi$ / $K$ separation, allowing for studying final states involving neutral 
particles like $\pi^0$ or $\eta$ as well as hidden strangeness, respectively. We discuss the status of 
ongoing analyses with specific focus on diffractively produced $(\pi^0\pi^0\pi)^{-}$
as well as $(K\bar{K}\pi)^{-}$ final states.
\end{abstract}

\ccode{PACS numbers: 13.25.-k,13.85.-t, 14.40.Be, 29.30.-h}

\section{Introduction}	
The existence of states beyond the constituent quark model (CQM) has been speculated about almost since the introduction of 
colour\cite{Jaffe:1976,Barnes:1983}. Due to the gluon self-coupling via colour-charge, so-called hybrid mesons, 
$q\bar{q}$ states with an admixture of gluons, and glueballs, states with no quark content, consisting of (constituent) 
gluons only, are allowed within Quantum Chromodynamics. While light glueballs would not be observable as pure states,
as they would mix with ordinary $q\bar{q}$-mesons, hybrid mesons ($q\bar{q}g^n$) are promising candidates to search 
for resonances beyond the CQM. Especially, as the lowest mass candidate is predicted\cite{Morningstar:2004} to have exotic 
quantum numbers of spin, parity and charge conjugation $J^{PC}=1^{-+}$ not attainable by ordinary $q\bar{q}$ states, and a mass 
between 1.3 and 2.2\,GeV/c$^2$.
Two experimental $1^{-+}$ hybrid candidates in the light-quark sector have been reported in the past in different decay 
channels, the $\pi_1(1400)$ and the $\pi_1(1600)$. In particular the resonant nature of the $\rho\pi$ decay channel of the 
latter is highly disputed\cite{Amelin:2005,Dzierba:2006}. COMPASS has started to shed new light on the 
puzzle by the observation of an exotic $1^{-+}$ signal in the 2004 data, consistent with the famous $\pi_1(1600)$; it shows 
a clean resonant behaviour\cite{Alekseev:2009a}.            
\newline
\noindent
The COMPASS spectrometer\cite{compass:2007} at the CERN SPS features electromagnetic calorimetry and a Ring Imaging Cherenkov detector. Photon detection in a wide angular range with high resolution is crucial for decay channels involving $\pi^{0}$, $\eta$ or $\eta'$, and the RICH allows for final state particle identification. 
A good separation of pions from kaons enables the study of kaonic final states. Not only production of 
strangeness with the pion beam can thus be studied but also kaon diffraction, using the incoming kaon beam. The COMPASS data 
taken in 2008/09 with a 190\,GeV/c $\pi/K$ or $p$ beam provide thus excellent opportunity for simultaneous observation of new states in various decay modes 
within the same experiment. Moreover, the data contain subsets with different beam projectiles and targets, 
allowing for systematic studies not only of diffractive but also central production and Primakoff reactions.  
\begin{figure}[tp!]
  \begin{minipage}[h]{.49\textwidth}
    \begin{center}
      \includegraphics[clip,trim= 3 0 22 9,width=0.9\linewidth,
	angle=0]{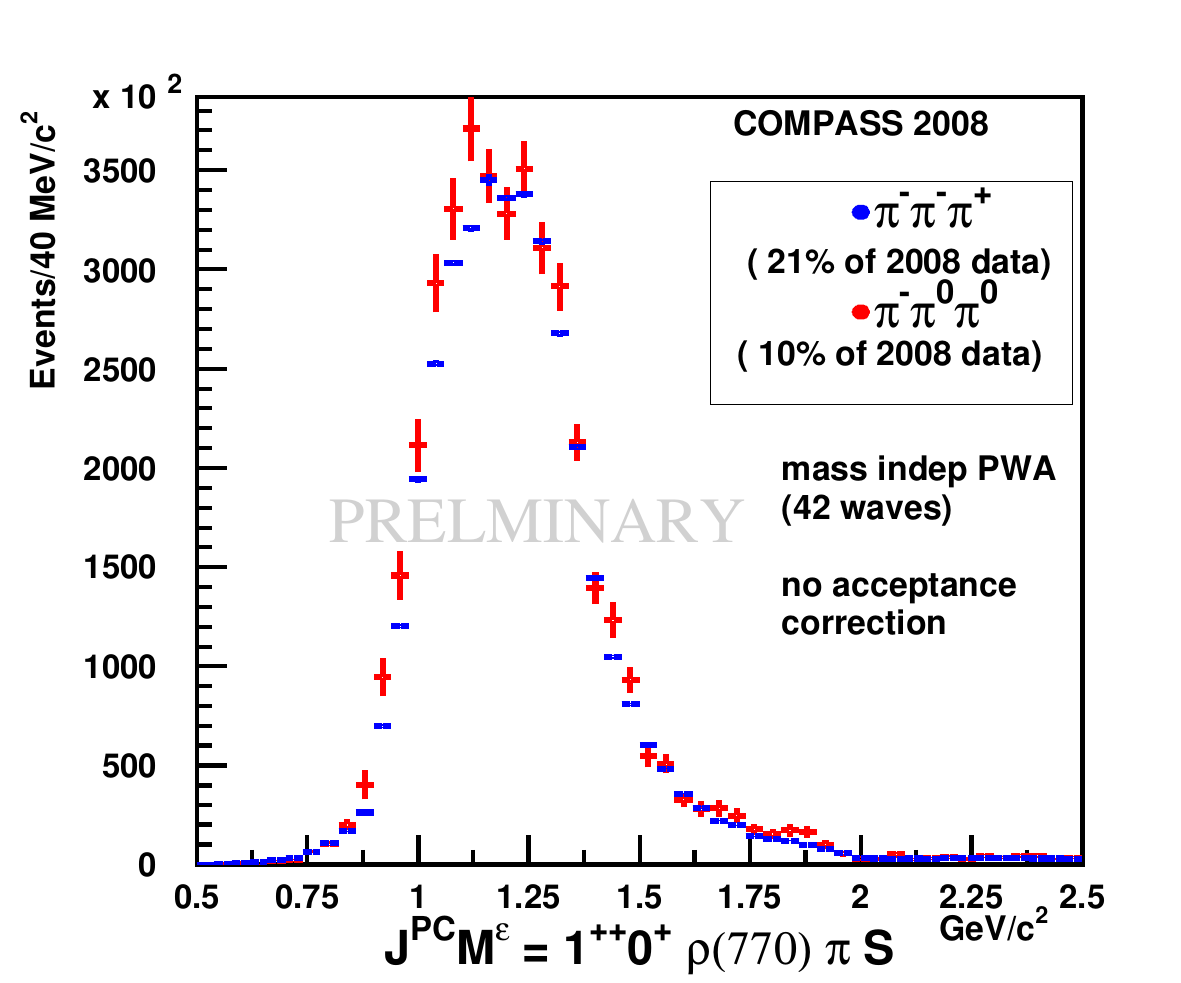}
    \end{center}
  \end{minipage}
  \hfill
  \begin{minipage}[h]{.49\textwidth}
    \begin{center}
      \includegraphics[clip,trim= 3 0 22 9,width=0.9\linewidth,
     angle=0]{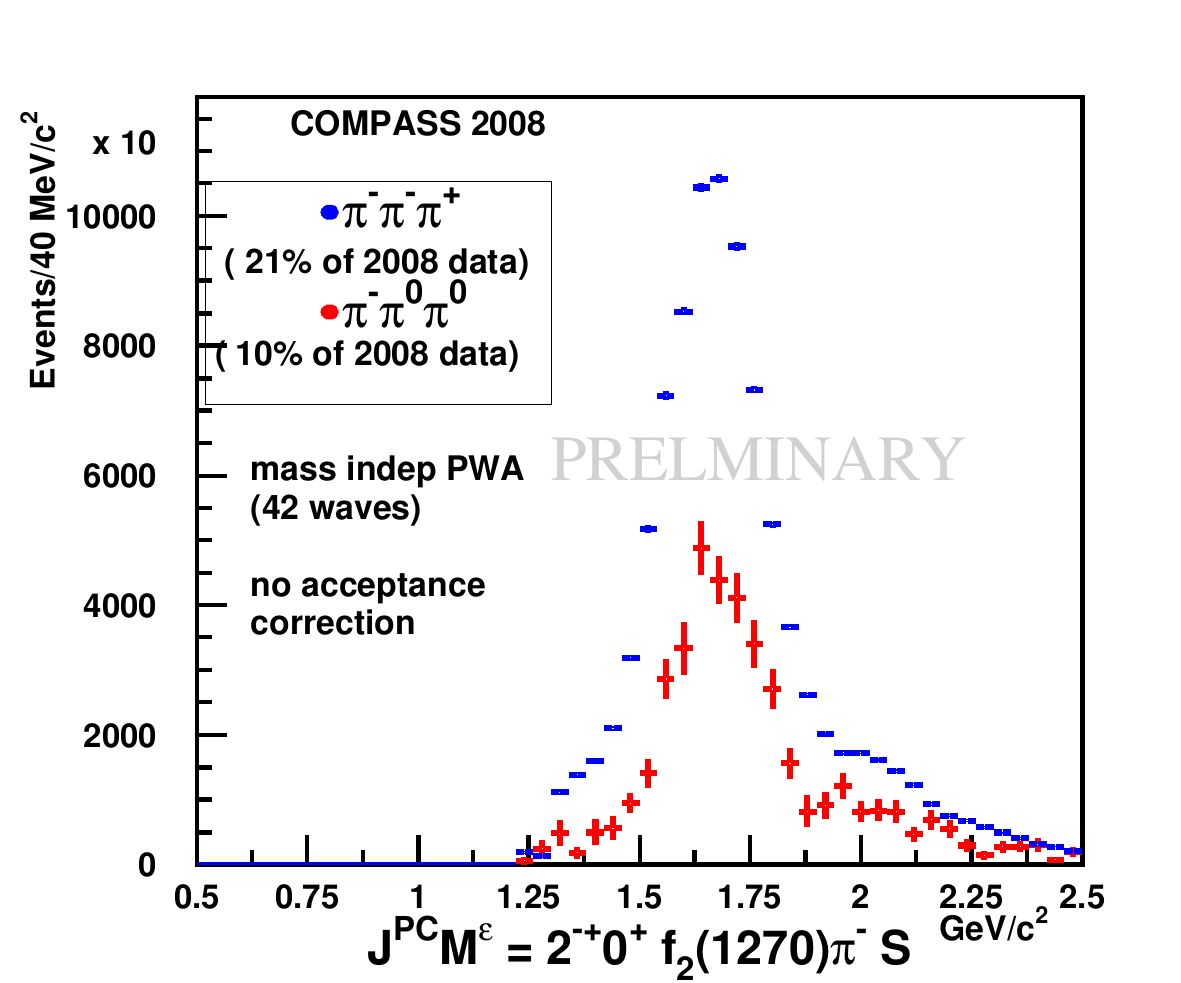}
    \end{center}
  \end{minipage}
      \caption{Comparison of 3$\pi$ analyses neutral vs. charged mode (2008 data): Exemplary intensities of main waves. For comparison, neutral and charge mode intensities are normalised to the well-established $a_{\rm 2}(1320)$ in the $2^{\rm ++} 1^{\rm +}[\rho\pi]D$ wave. 
      {\it Left:} $1^{\rm ++} 0^{\rm +}[\rho\pi]S$ wave ($a_{\rm 1}(1260)$): .
      {\it Right:} $2^{\rm -+} 0^{\rm +}[f_{\rm 2}(1270)\pi]S$ wave  ($\pi_{\rm 2}(1670)$).}
\label{fig:PWA2008}
\end{figure}
\section{First results on neutral and kaonic channels}
An important cross-check of all analyses is the test for isospin symmetry in the observed spectra.
The $\rho\pi$ decay channel of the $\pi_1(1600)$ for example, can be studied in two modes of 3$\pi$ 
final states, $\pi^{-}\pi^{+}\pi^{-}$ (charged) and $\pi^{-}\pi^{0}\pi^{0}$ (neutral), respectively.
Depending on the underlying isobars, the relative contribution should follow isospin conservation.
This is shown in Fig.\,\ref{fig:PWA2008}. A first, preliminary partial-wave analysis (PWA) of main waves in 
diffractively produced 3$\pi$ events has been performed for both modes, for details see\cite{nerling:2009}. 
To compensate different statistics, the wave 
intensities shown are normalised to the well-known $\rho\pi$ decay of the $a_2(1320)$ to make them comparable. 
We find similar intensities for the $\rho\pi$ decay, whereas a suppression factor of two is observed for 
the wave decaying into $f_2\pi$ as expected due to the Clebsch-Gordon coefficients.
Further ongoing analyses involving neutrals cover $\pi^{-}\eta$, $\pi^{-}\eta\eta$ (search for the 
$\pi_1(1400)$ and lightest $0^{++}$ glueball candidate) as well as $\pi^{-}\pi^{-}\pi^{+}\pi^{0}$,
$\pi^{-}\pi^{-}\pi^{+}\eta$ and $\pi^{-}\pi^{-}\pi^{+}\pi^{0}\pi^{0}$ final states (accessible isobars: 
$f_1, b_1, \eta, \eta', \omega$), for which COMPASS has recorded significantly higher statistics with respect 
to previous experiments, covering all spin-exotic meson decay channels in the light quark sector reported 
in the past.     
\begin{figure}[tp!]
  \begin{minipage}[h]{.49\textwidth}
    \begin{center}
      \includegraphics[clip,trim= 10 5 25 15,width=1.0\linewidth,
	angle=0]{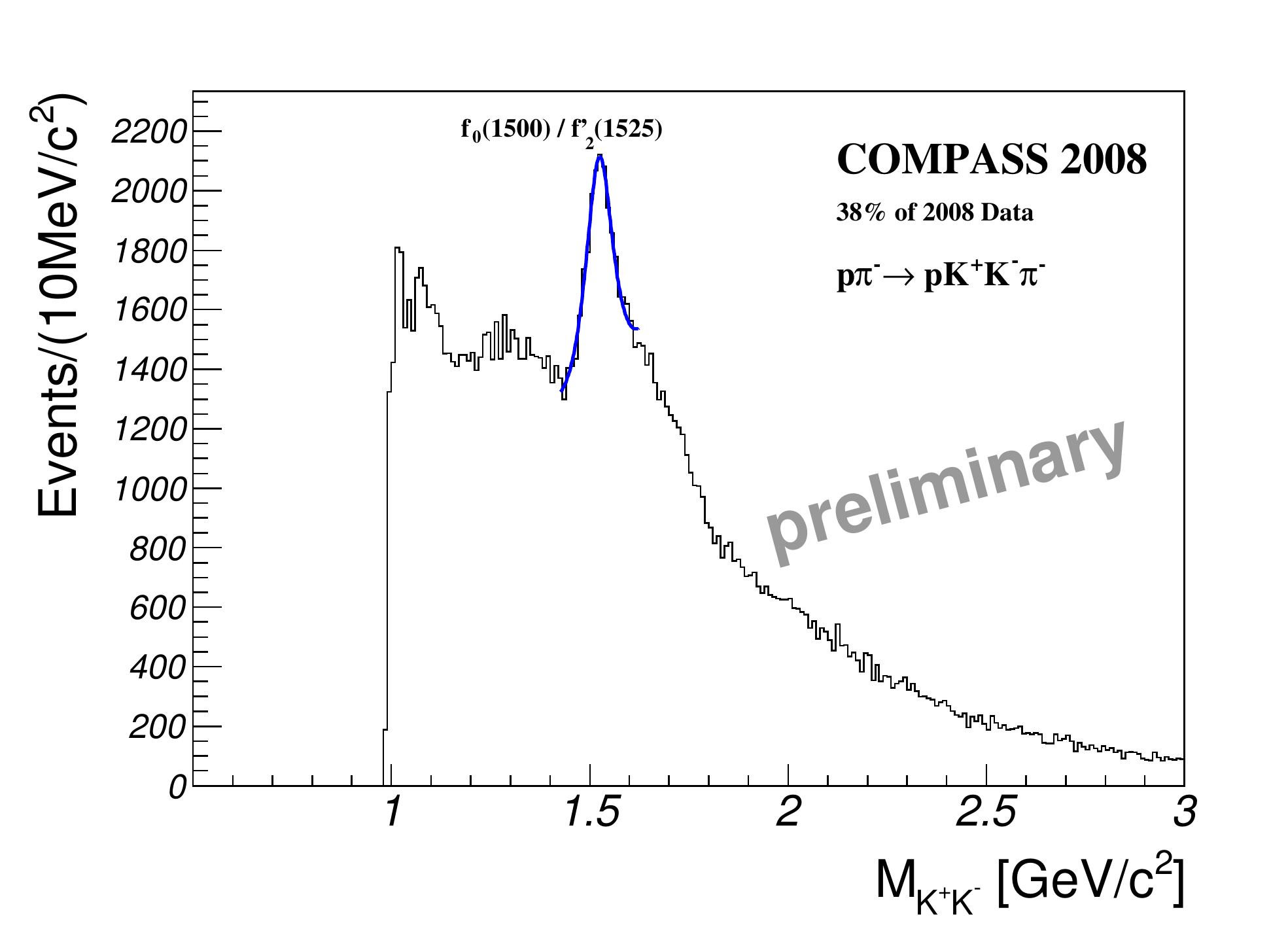}
    \end{center}
  \end{minipage}
  \hfill
  \begin{minipage}[h]{.49\textwidth}
    \begin{center}
      \includegraphics[clip,trim= 10 0 25 20,width=1.0\linewidth,
     angle=0]{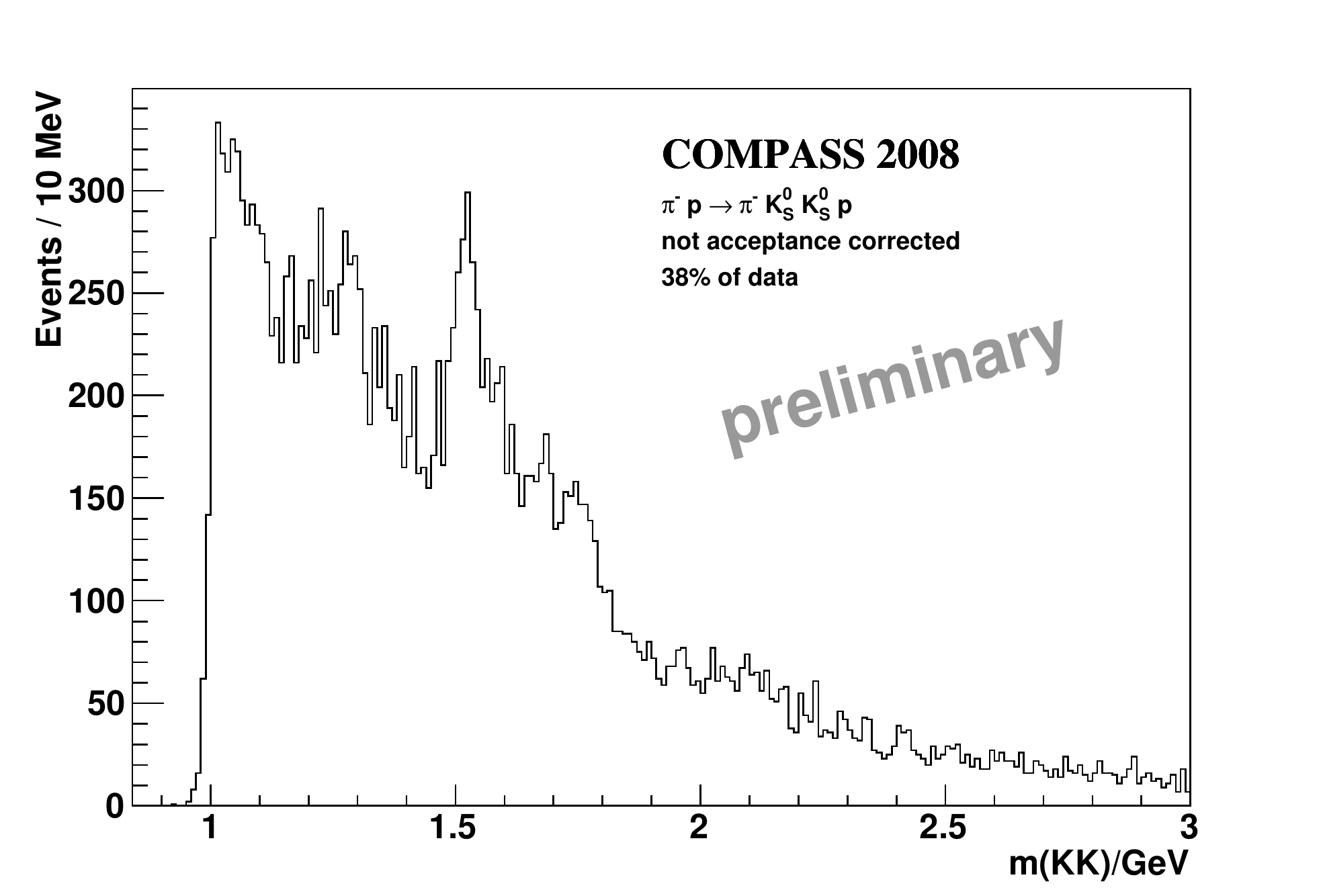}
    \end{center}
  \end{minipage}
      \caption{Invariant mass spectra of $(K\bar{K}\pi)^{-}$ systems: 
	{\it (Left)} $K^+K^-$ (with $p_{K^-} \le 30$\,GeV/c) {\it (Right)} $K^0_sK^0_s$.}
      \label{fig:Kaons}
\end{figure}
\newline
\noindent
Final states with strangeness are interesting for both, the glueball search in central production 
as well as diffractive production of hybrids. Exemplary we show the $K\bar{K}$ subsystems out of the 
$(K\bar{K}\pi)^{-}$ system again for two modes, $K^{+}K^{-}\pi^{-}$ and $K^{0}_s K^{0}_s\pi^{-}$ final 
states, respectively. Both spectra show a clear structure around the expected $f_0(1500)$, for details see\cite{tobi:2009}. Further ongoing 
analyses cover the $(K\bar{K}\pi)^{0}$ system as well as kaon diffraction into $K^{-}\pi^{+}\pi^{-}$ final 
states (PWA under preparation). 
\section{Summary \& conclusions}
COMPASS has collected data with high-intensity hadron beams ($\pi^{\pm},K^{\pm},p$) on nuclear and liquid 
hydrogen targets. The newly taken data sample exceeds the world data by a factor of 10-100, allowing to address 
open issues in light-mesons spectroscopy at good accuracy, even in the mass region beyond 2\,GeV/c$^2$.   

\section*{Acknowledgements}
This work is supported by the BMBF (Germany), particularly the ``Nutzungsinitiative CERN''.

\end{document}